\documentclass[sn-mathphys, numbered]{sn-jnl}

\usepackage{graphicx}%
\usepackage{multirow}%
\usepackage{amsmath,amssymb,amsfonts}%
\usepackage{amsthm}%
\usepackage{mathrsfs}%
\usepackage[utf8]{inputenc}
\usepackage[T1]{fontenc}
\usepackage[title]{appendix}%
\usepackage{xcolor}%
\usepackage{textcomp}%
\usepackage{manyfoot}%
\usepackage{booktabs}%
\usepackage{siunitx}

\begin{document}

\title{The miniSLR: A low-budget, high-performance satellite laser ranging ground station}

\author*[1,2]{\fnm{Daniel} \sur{Hampf}}\email{daniel.hampf@digos.eu}

\author[1]{\fnm{Felicitas} \sur{Niebler}} 

\author[1]{\fnm{Tristan} \sur{Meyer}} 

\author[1]{\fnm{Wolfgang} \sur{Riede}} 

\affil*[1]{\orgdiv{Institute of Technical Physics}, \orgname{German Aerospace Center}, \orgaddress{\street{Pfaffenwaldring 38-40}, \city{Stuttgart}, \postcode{70569}, \state{BW}, \country{Germany}}}

\affil[2]{\orgname{DiGOS Potsdam GmbH}, \orgaddress{\street{Telegrafenberg}, \city{Potsdam}, \postcode{14473}, \state{BB}, \country{Germany}}}


\abstract{Satellite Laser Ranging (SLR) is an established technique providing very accurate position measurements of satellites in Earth orbit. However, despite decades of development, it remains a complex and expensive technology, which impedes its further growth to new applications and users.

The miniSLR implements a complete SLR system within a small, transportable enclosure. Through this design, costs of ownership can be reduced significantly, and the process of establishing a new SLR site is greatly simplified. A number of novel technical solutions have been implemented to achieve a good laser ranging performance despite the small size and simplified design.

Data from the initial six months of test operation has been used to generate a first estimation of the system performance. The data includes measurements to many of the important SLR satellites, such as Lageos, Etalon and most of the geodetic and Earth observation missions in LEO. It is shown that the miniSLR achieves sub-centimetre accuracy, comparable with conventional SLR systems.

The miniSLR is an engineering station in the International Laser Ranging Service (ILRS), and supplies data to the community. Continuous efforts are undertaken to further improve the system operation and stability.}




\maketitle

\section{Introduction}

Satellite laser ranging (SLR) is a powerful tool for geodesy, mission support and fundamental science (\cite{pearlman_SLR_intro}). In the future, it may also be used for space situational awareness, precise orbit determination and conjunction assessment (\cite{hampf_2021}). However, the effort to construct and operate an SLR ground station is considerable, and poses an entry barrier for new users and applications. Thus, the existing SLR network (see International Laser Ranging Service, ILRS (\cite{ILRS_web}, \cite{ILRS_paper})), still suffers from significant gaps in global coverage, especially in the Global South (\cite{otsubo_2016}). But also elsewhere, the demand for new SLR stations is growing: An increasing number of SLR missions causes a high load on existing stations, which can only be met with a tight schedule and in favourable weather conditions. More stations may be needed in the future to satisfy all ranging requirements. Furthermore, some older stations reach the end of their lifetime and will need replacement in coming years (\cite{Wilkinson_2018}). 

On the other hand, technical advances of the last twenty years have opened the possibility to construct much smaller, simpler and cheaper SLR ground stations. Compact and powerful lasers, better detectors, faster readout electronics and PCs, and inexpensive but precise direct drive telescope mounts are key technologies for this development. The goal of the miniSLR project has been to combine these novel technologies for the first time into a working prototype of a new generation of SLR ground station. At a fraction of the cost of a conventional SLR system, it is designed to reach the same performance in terms of precision, stability and tracking capabilities. 

A first version of the miniSLR has been constructed and set up at the DLR (German Aerospace Center) in Stuttgart. In its current configuration, it commenced experimental operation in November 2022. It has been accepted into the ILRS as engineering station, and data from measured passes have been uploaded to the European Data Center (\cite{EDC_web}). Table \ref{tab:station_parameters} lists the station's ILRS IDs and coordinates.

\begin{table}
	\begin{tabular}{ll}
		\toprule
		System name & miniSLR \\
		4-Character Code & SMIL \\
     	CDP System Number & 52 \\
		CDP Occupation Number & 02 \\
		IERS DOMES Number & 10916S001 \\
		CDP Pad ID & 7816 \\
		Location & Stuttgart, Germany \\
		Latitude & 48.748893981739° N \\
		Longitude & 9.102599520211° E \\
		Elevation & 533.240 m \\
		\bottomrule
	\end{tabular}
	\caption{ILRS station parameters. The coordinates are approximate, based on a GNSS survey.}
	\label{tab:station_parameters}
\end{table}

This paper describes the technical design (Section \ref{sec:design}) and results from the first six months of test operation (Section \ref{sec:results}). In the final Section \ref{sec:conclusion}, an outlook to the next steps in development, and to the potential impact of this new development for the SLR and space geodesy community is given.

\section{System set-up}
\label{sec:design}

\subsection{Design innovations}
\label{sec:design_approach}

This section highlights the main design features of the miniSLR that enable the reduction of size and complexity of the system. A full system overview is given in subsequent Section \ref{sec:hardware_setup}.

\subsubsection{Transportability and small overall size}
\begin{figure}[t]
	\centering
	\includegraphics[width=0.5\textwidth]{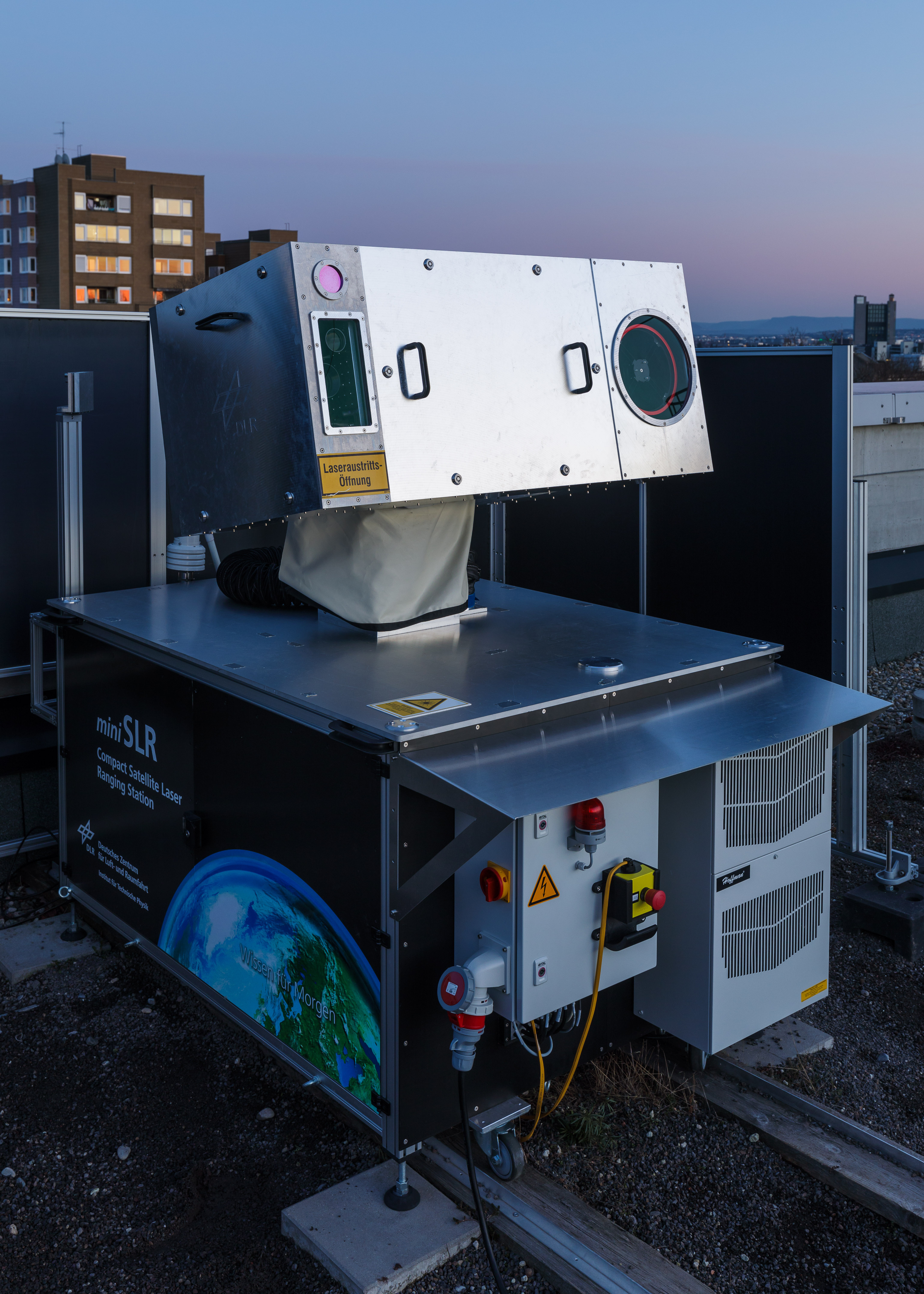}
	\caption{The miniSLR prototype on the roof of the DLR institute building. The enclosure in the bottom contains most of the electronics and IT. The top compartments house the receive and transmit telescopes, the laser head, cameras, detectors and beam control optics.}
	\label{fig:miniSLR_pic}
\end{figure}

The main design goal of the miniSLR has been a significant reduction in the size of the system. Using a receive telescope of only \SI{20}{\centi\metre} aperture and a small direct drive astronomy mount, the whole system could be integrated within and on top of an aluminium enclosure with a footprint of 130 x 230 cm (see Figure \ref{fig:miniSLR_pic}). At a weight of about \SI{600}{\kilo\gram}, the enclosure can be moved around on wheels and installed for operation on any flat and stable surface. 

The main advantages of this integrated design are:
\begin{itemize}
	\item Lower production cost
	\item Lower maintenance cost
	\item The system can be integrated and validated at factory, before it is moved to its operation site, thus lowering the effort for installation and decreasing delays in the commissioning process.
	\item No need for civil engineering, building permits and construction works, thus significantly lowering costs, effort and time needed for installation.
	\item Re-location can be achieved easily, if necessary.
\end{itemize}

Some of these advantages have already been demonstrated by the French Transportable Laser Ranging Station (FTLRS, \cite{ftlrs_2001}).

\subsubsection{Fully sealed, no dome}
The miniSLR system is fully sealed, thus avoiding the need for a dome with movable parts. This offers two main advantages: First, in case of a catastrophic failure (e.g. complete power loss, mechanical blockage etc.), the system remains in an inherently safe status (i.e. protected from rain). Recovery and repair can be planned and conducted with much less urgency than in the case of a dome that can no longer be closed. Second, the whole system is air-conditioned and retains a constant temperature in all parts. Combined with the relatively short cable lengths, this increases the stability of the timing measurement.

\subsubsection{High repetition rate Q-switch laser}
Due to the small aperture of the receive telescope, a rather high power laser is needed to achieve sufficient returns. On the other hand, relatively stringent size and weight limitations apply, since the laser head must be mounted in the top compartment. This was resolved by using high-repetition laser ranging (\cite{hampf_2019}) with a small Q-switched diode laser. In this context, it offers three advantages:
\begin{itemize}
	\item Sufficient average power at a very small footprint: At a size of \SI{12}{\centi\metre} x \SI{8}{\centi\metre} x \SI{4}{\centi\metre}, the laser offers a power of 5 Watts (\SI{100}{\micro\joule} at \SI{50}{\kilo\hertz}).
	\item Due to the low pulse energy, single photon operation is inherently ensured without the need for additional attenuation components, thus further simplifying the design.
	\item The high repetition rate results in a high number of returns for most targets, which decreases the statistical error of the average data points. This enables sufficiently precise measurements despite the relatively long pulse duration of \SI{500}{\pico\second} (FWHM).
\end{itemize}

\subsubsection{Laser Ranging at 1064 nm (near-infrared)}
Using the Nd:YAG fundamental wavelength of \SI{1064}{\nano\metre} for laser ranging has been discussed for many years (e.g.\ \cite{voelker_2013}), and has recently been implemented in a number of SLR systems (\cite{courde_2017}, \cite{xue_2016}, \cite{eckl_2017}, IZN-1 at Teneriffa). Nevertheless, most systems still use frequency doubling to obtain green laser light at \SI{532}{\nano\metre}. This choice is primarily due to the available receive detectors: Up until a few years ago, single photon detectors with picosecond timing precision have only been realized for the visible light spectrum (either photomultiplier tubes or silicon based geiger mode avalanche photo diodes). Today, however, InGaAs SPADs (single photon avalanche diodes based on indium gallium arsenide) achieve sufficient timing precision and good sensitivity at \SI{1064}{\nano\metre}. With these, ranging at the fundamental Nd:YAG wavelength becomes more favourable for a number of reasons:
\begin{itemize}
	\item Avoiding conversion losses of about a factor of four, thus generating more usable power with a same laser (important to keep laser size and weight small)
	\item Avoiding complexity of additional frequency doubling optics
	\item Slightly better atmospheric transmission
	\item Less noise from sky brightness, especially at daylight (blue sky)
\end{itemize}

\subsection{Hardware set-up}
\label{sec:hardware_setup}

\begin{figure}[pt]
	\centering
	\includegraphics[width=0.9\textwidth]{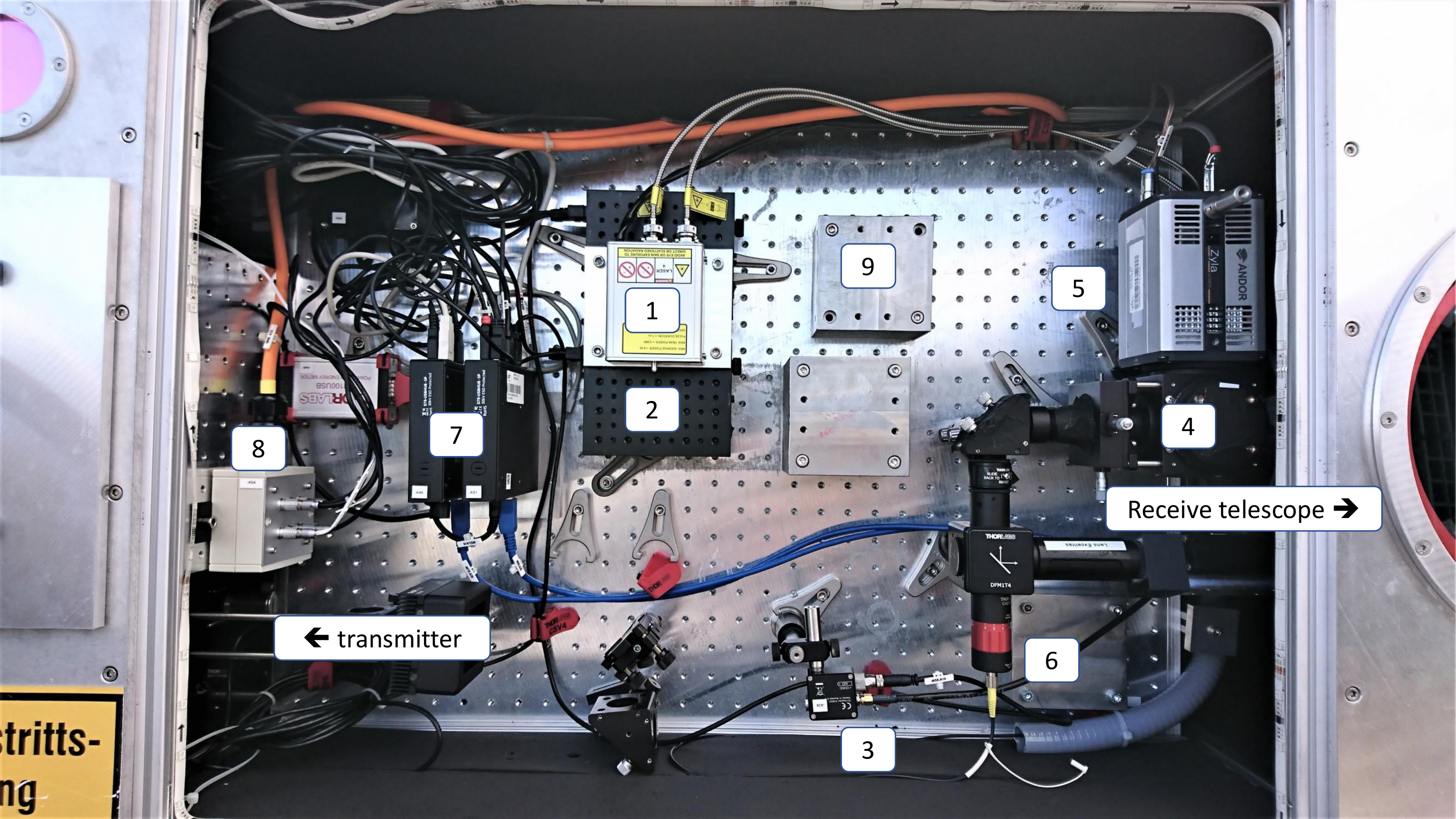}
	\caption{The central compartment of the miniSLR head: (1) Laser head; (2) Thermo-electric elements (TEC) for laser temperature control; (3) start photodiode; (4) dichroic beam splitter; (5) tracking camera; (6) fibre coupling for single photon detector; (7) programmable USB hubs; (8) 12 V power distribution; (9) counter weights to balance mount elevation axis.}
	\label{fig:central_board}
\end{figure}

\begin{figure}[pt]
	\centering
	\includegraphics[width=0.8\textwidth]{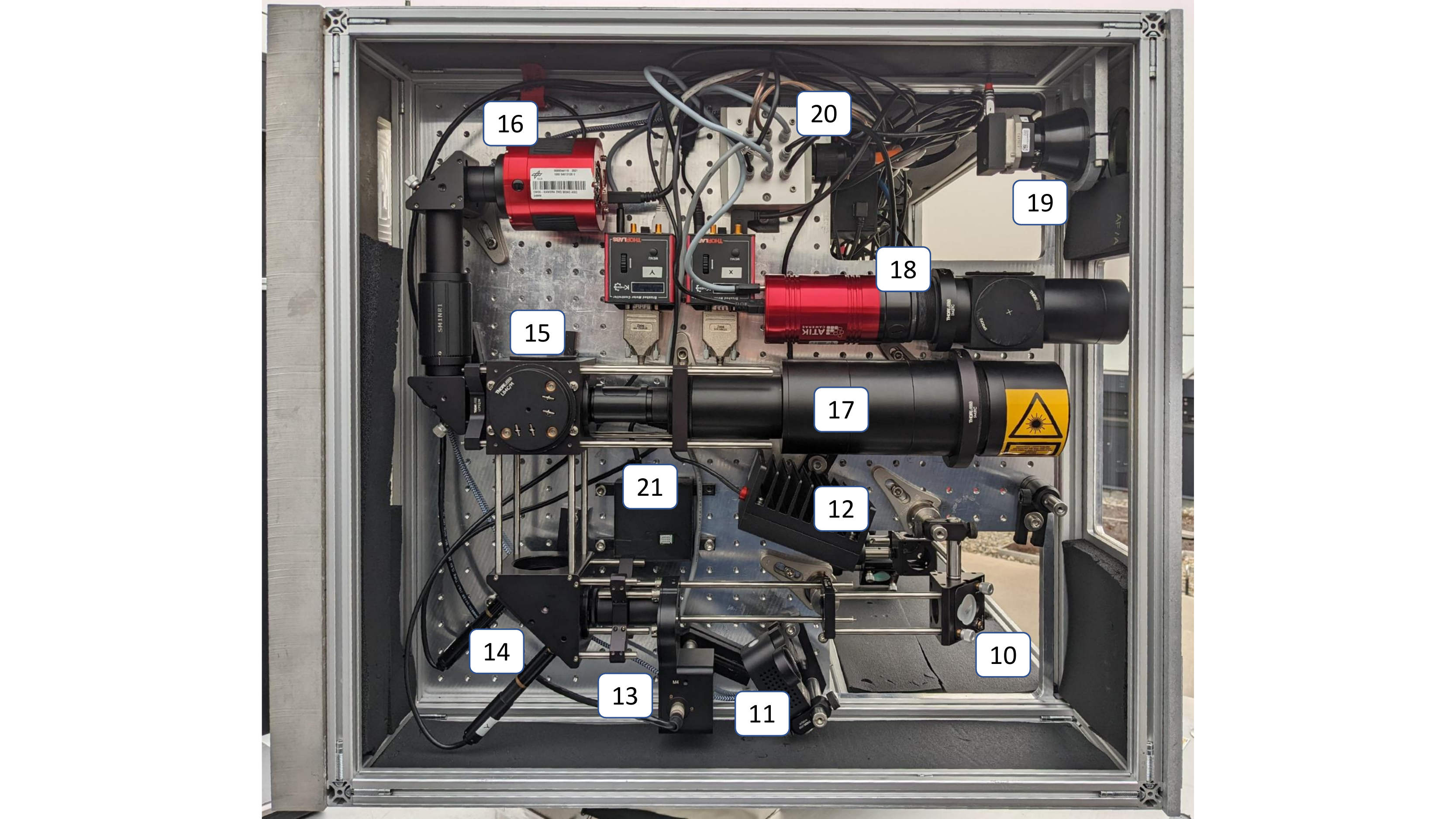}
	\caption{The transmitter compartment of the miniSLR head: (10) incoming beam from central compartment; (11) attenuator flip mirror; (12) laser power meter; (13) safety shutter; (14) motorized beam steering mirror; (15) dichroic mirror guiding \SI{1064}{\nano\metre} towards exit aperture; (16) transmitter camera; (17) beam expander; (18) backscatter camera; (19) aircraft camera; (20) power distribution; (21) thermometer / hygrometer}
	\label{fig:tx_box}
\end{figure}

This section gives a brief overview of the hardware set-up of the miniSLR. References to "item NN" relate to the indicators in Figures \ref{fig:central_board} and \ref{fig:tx_box}.


\subsubsection{Tracking and mechanics}
Tracking is realized using an Astelco NTM-600 direct drive mount. It allows programming of custom trajectories, which are followed with a high timing precision owing to an internal GNSS clock. Thus, sufficiently accurate tracking to satellites with good predictions can be achieved. For simplicity, the pointing model is done by the main control software (see Section \ref{sec:software}) rather than the mount's own firmware. For rain protection, the mount is wrapped in a Telegizmos telescope cover.

The optical set-up, including receive and transmit telescopes, is mounted on three optical breadboards installed on top of the mount. This enables a high degree of flexibility in the optical configuration, which is of paramount importance for an experimental prototype.

\subsubsection{Transmit path}
Laser pulses are produced by a Standa MOPA-4 diode laser (item 1). Its specifications are ideally suited for a small SLR system: The tiny laser head can easily be installed on the moving platform. A pulse duration of \SI{500}{\pico\second} FWHM is sufficiently short to achieve a high precision in averaged normal point data (see Sections \ref{sec:precision_theo} and \ref{sec:NPT_precision_theo} for a more detailed discussion on ranging precision). A pulse energy of \SI{85}{\micro\joule} is enough to achieve returns from all relevant satellites, and a repetition rate of \SI{50}{\kilo\hertz} provides a high amount of data points for effective averaging (see also Section \ref{sec:exp_return_rates}).

Temperature control for the laser head is realized by two Thorlabs PTC1/M thermoelectric elements. They keep the laser head at constant 22° Celsius and can dissipate up to \SI{35}{\watt} of heat (item 2).

Pulse emission times are recorded with a standard photodiode (Thorlabs DET08C/M) using a fraction of light leaking through the first mirror (item 3). From the mirror, the light is guided towards the transmitter compartment (item 10).

For calibration, laser power needs to be strongly attenuated in order to not saturate the detector. This is achieved by a flip mirror carrying a reflective neutral density or laser line filter (item 11). It is closed by default and only opens when the system is tracking a satellite. While closed, the laser power is directed into a power meter (item 12). The laser average power is thus monitored and recorded each time a calibration run is performed. The subsequent safety shutter (item 13) is also closed by default and opens only for ranging measurements. It is spring loaded and requires a regular "open" signal from the software to open and remain open (for more information on the safety system, see also Section \ref{sec:safety}).

The motorized beam mirror (item 14) is needed to fine-control the laser beam direction in relation to the main system pointing. It is moved by two Thorlabs Z806 motors, controlled by the main software. 

The dichroic mirror (item 15) guides the infrared laser light towards the exit aperture, while incoming light passes through to the transmitter camera (item 16). This camera records the field of view seen by the beam expander and can be used during the initial alignment of the system. The beam expander itself (item 17) is a simple Galileo type telescope with a one-inch negative and a three-inch positive lens. It increases the beam diameter by about a factor of five, thus decreasing the beam divergence and improving the beam pointing resolution.

The backscatter camera (item 18) is used to monitor the laser beam in the atmosphere, and fine-align it towards the target. 

\subsubsection{Receive path}

The receive path starts with a \SI{20}{\centi\metre} aperture Newton telescope (ASA H8). A dichroic mirror on its exit port (item 4) guides visible light towards the main tracking camera (item 5), while transmitting the returning infrared laser light towards the single photon receiver. Two bandpass filters block light from \SI{900}{\nano\metre} to \SI{1700}{\nano\metre}, with a \SI{1}{\nano\metre} wide window at \SI{1064}{\nano\metre}. While not really required at night, these filters are essential for daylight ranging. For simplicity, they are permanently installed and not removed for night time ranging.

The light is coupled into a \SI{105}{\micro\meter} multi-mode optical fibre connected to an Aurea SPD-OEM-NIR single photon detector, which generates the stop signal for the ranging measurement. 

Table \ref{tab:minislr_specs} summarizes the specifications of the optical system (transmit and receive). These values are used in Section \ref{sec:exp_return_rates} to calculate the expected photon return rates.


\begin{table}
	\begin{tabular}{rrl}
		\toprule
		Transmit aperture  & \SI{7.5}{\centi\metre} & \\
		Beam diameter     & \SI{5}{\centi\metre} & \\
		Receive aperture (nominal) & \SI{20}{\centi\metre} & \\
		Obscuration & 25\% & due to secondary mirror in telescope \\
		Laser pulse energy & \SI{85}{\micro\joule} & (measured) \\
		Laser repetition rate & \SI{50}{\kilo\hertz} & (measured) \\
		Operating wavelength & \SI{1064}{\nano\metre} & \\
		Beam divergence & \SI{50}{\micro rad} & (half angle, estimated) \\
		Beam stability & \SI{50}{\micro rad} & (half angle, estimated) \\
		Transmitter efficiency & 0.6 & (measured) \\
		Receiver efficiency & 0.1 & (estimated, losses e.g.\ in band-pass filter) \\
		Efficiency of detector & 30\% & (given by manufacturer) \\
		\bottomrule
	\end{tabular}	
	\caption{Main specifications of the miniSLR optical system. These values are also used for calculation of the expected return rates (Section \ref{sec:exp_return_rates}).}
	\label{tab:minislr_specs}
\end{table}

\subsubsection{Timing measurement and control}
System-wide frequency and time synchronisation is based on a Meinberg GPS-180 GNSS disciplined atomic clock, which provides a \SI{10}{\mega\hertz} sine wave, a 1 PPS signal, and the datum over serial interface.

The timestamps are recorded using a Swabian instruments Time Tagger Ultra with a nominal timing precision of \SI{9}{\pico\second}. Additionally to the start, stop and PPS signals, laser trigger and detector gate are recorded as well on separate channels for debugging and monitoring.

\begin{figure}[t]
	\centering
	\includegraphics[width=0.8\textwidth]{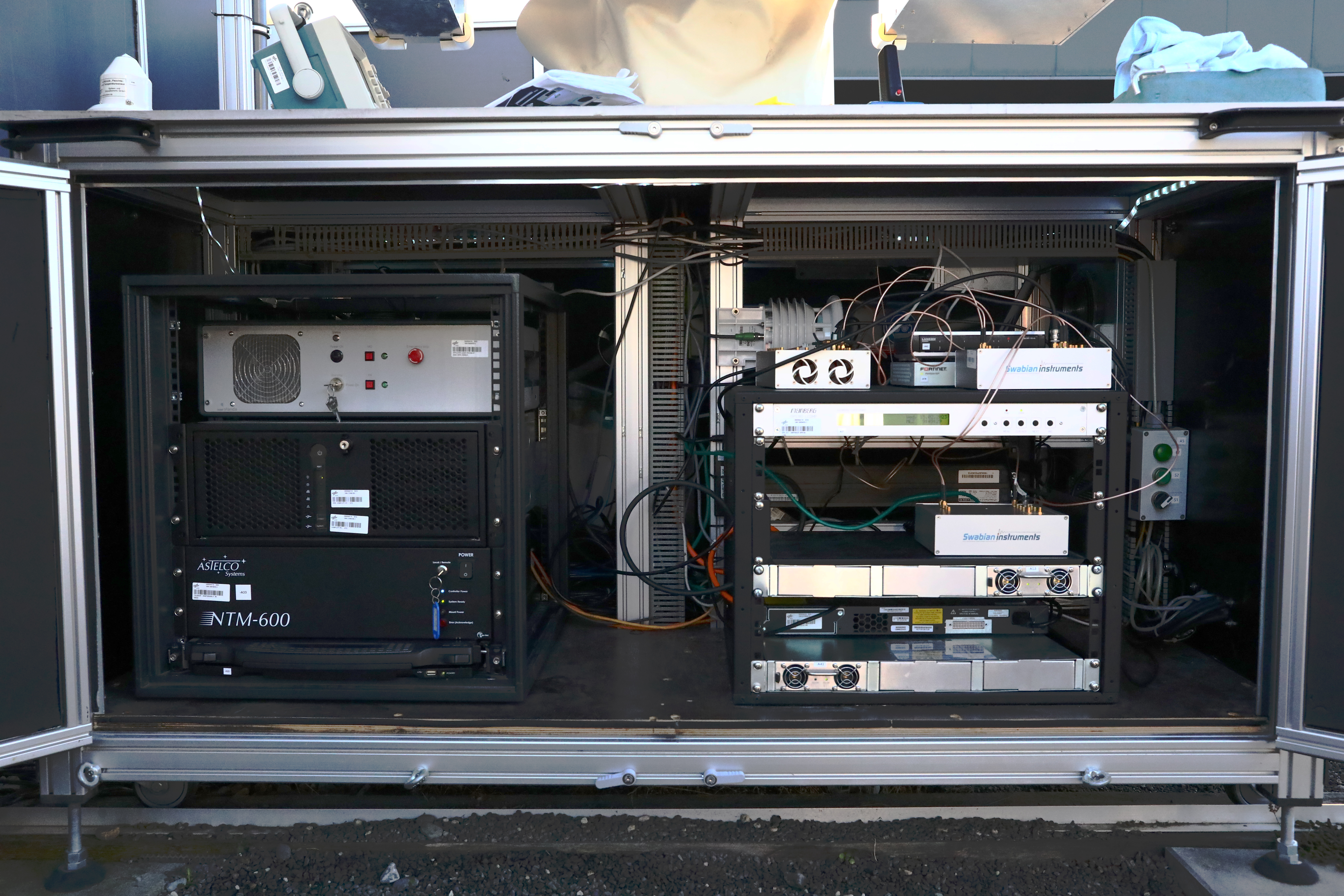}
	\caption{The inside of the cabinet houses the complete electronics for the system. Left rack top to bottom: Laser controller, PC, mount controller. Right rack: Internet router, trigger and gate controller (Swabian Instruments), atomic clock (Meinberg), event timer (Swabian Instruments), network switch, two \SI{12}{\volt} power supplies.}
	\label{fig:cabinet}
\end{figure}

Laser triggers and detector gate signals are generated by two Swabian Instruments Pulse Streamers. While one pulse generator runs a steady PPS-aligned \SI{50}{\kilo\hertz} trigger sequence for the laser, the other one produces a dynamically calculated gating sequence for the single photon detector, based on the expected time of flight to the target. Figure \ref{fig:cabinet} shows the electronics installed in the cabinet.

Timing calibration is done roughly once per hour during regular operation. For this, the attenuated laser beam is guided directly towards the receiver aperture via a 45° mirror and a 45° diffuse surface. The nominal range to the calibration target is given by the distance from the mount axes intersection to the virtual intersection of the two 45° surfaces, and measured to \SI{1.504}{\metre}.

\subsubsection{Aircraft and laser safety}
\label{sec:safety}
The output beam power is significantly above the maximum permissible exposure (MPE) defined in the laser safety norm EN\ 60825-1:2014. The miniSLR is thus classified as a class 4 laser system. While the beam expander reduces the power and energy density enough to avoid skin burns outside of the device, the limit for eye injuries is exceeded by about a factor of 200 at the exit aperture. Assuming a diffraction limited beam divergence of about \SI{50}{\micro rad}, the laser beam becomes eye-safe at a distance of about \SI{10}{\kilo\metre}.

Laser emission is automatically shut off by the attenuator and safety shutter, unless a "clear" signal is given from the following checks (conducted by a special module in the control software):
\begin{itemize}
	\item Pointing must be above the azimuth-dependent minimum elevation, which mirrors the adjacent buildings and obstacles. Below the elevation mask, the safety shutter can be opened, but only if the eye-safe attenuation is activated. This is mainly used for time calibration.
	\item Telescope must be tracking a target, not slewing or idling. The target must be whitelisted for laser ranging.
	\item No aircraft must be within 20° of the beam, within a distance of up to \SI{20}{\kilo\metre}. Aircraft positions are received by a data stream from the German Air Traffic Authorities. For cross-check, a local ADS-B receiver (Jetvision Radarcape) and a thermal infrared camera (FLIR Tau-2, item 19) are installed.
	\item Operation status code must be nominal for a number of critical components, such as shutter and attenuator. Telescope pointing information must be up to date.
\end{itemize}

To ensure workplace safety, warning lamps, emergency stop button, access control and laser hazard signs are installed.

\subsubsection{Slow control}
To facilitate remote operation and simple trouble shooting, most power lines can be switched independently by software. This can be used e.g.\ to remotely restart components that are not working nominally. Programmable USB Hubs (Acroname 3P) are used to connect devices to the PC, which allow detailed monitoring of each USB port (data, power) and power-cycling USB-powered devices by software.

Temperature, humidity and air pressure are continuously recorded inside and outside of the device for monitoring, safety and SLR data evaluation.

\subsection{Software}
\label{sec:software}

\begin{figure}[t]
	\centering
	\includegraphics[width=1\textwidth]{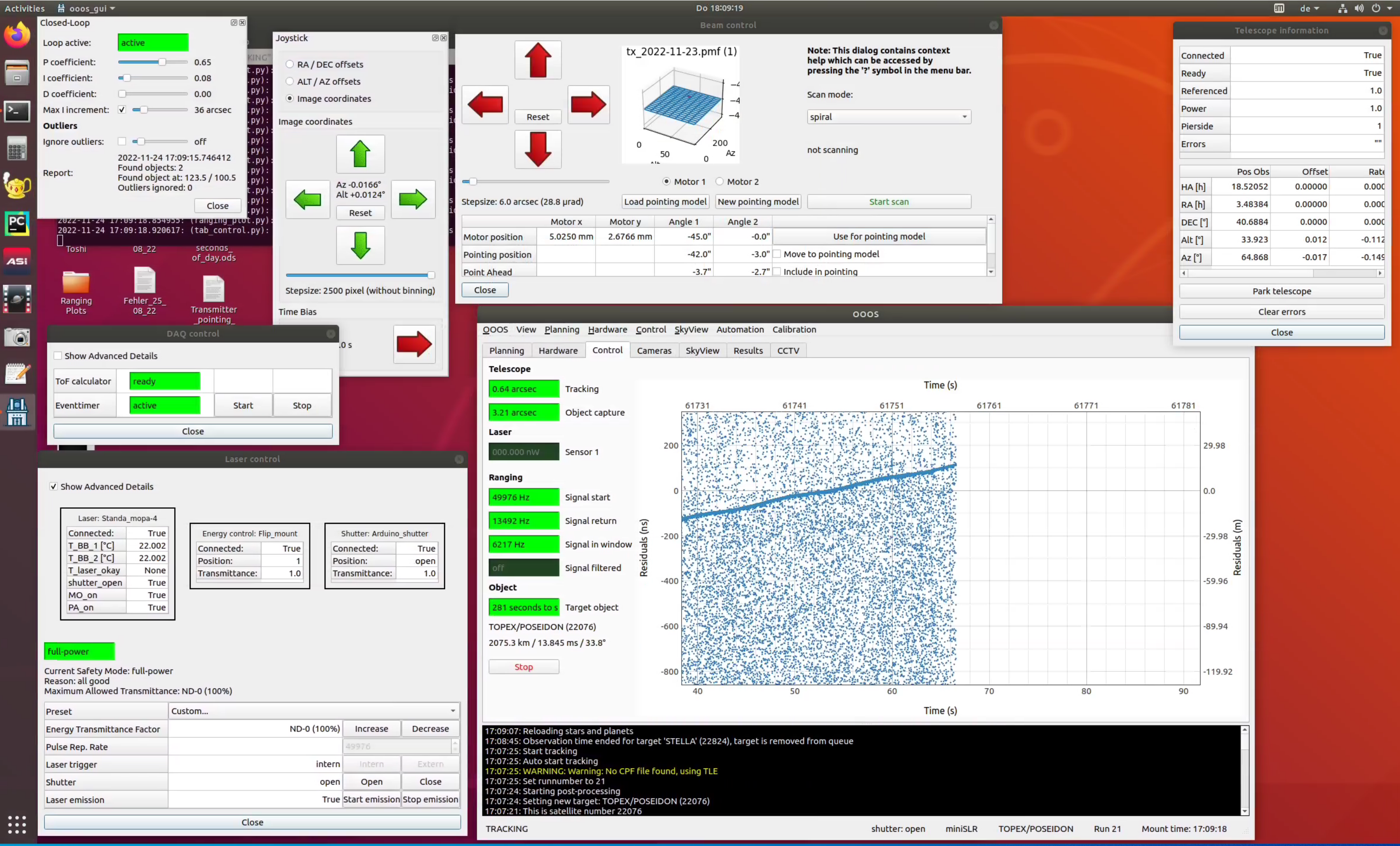}
	\caption{Screenshot of OOOS during a ranging observation of satellite TOPEX.}
	\label{fig:screenshot}
\end{figure}

To operate the miniSLR, the control software OOOS (orbital objects observation software), developed at the institute for the previous laser ranging station "Uhlandshöhe", has been refined and improved. It is written almost entirely in python to facilitate rapid development and easy debugging. Exploiting multiprocessing and fast computing libraries (e.g. numpy), the software can handle all control tasks in real-time on a standard Linux or Windows PC. Special focus has been put into designing a clear and tidy graphical user interface (GUI), thus enabling fast training of observers and efficient work. A number of automation functions take over most of the standard tasks, however a complete "hands-off" operation has not yet been achieved for laser ranging. 

A range of processing nodes (also called daemons) take over different blocks of control. The nodes are loosely coupled to each other over a TCP/IP protocol. The GUI acts as a central node, orchestrating the work of the daemon nodes and channelling all user interaction (input and output). The daemon nodes connect to an abstract hardware layer, which in turn implements the actual device interfaces based on current configuration settings. Thus, changes to the system's hardware can easily be incorporated into the software.

Figure \ref{fig:screenshot} shows a screenshot of OOOS running during an actual SLR observation.

\subsection{Expected performance}
\subsubsection{Single shot precision}
\label{sec:precision_theo}
The timing uncertainty of a single range measurement ("single shot precision") is given by the timing uncertainties of all contributing components:

\begin{equation}\label{eq:precision_terms}
	\sigma_{\mathrm{single}} = \sqrt{\sigma^{2}_{\mathrm{L}} +\sigma^{2}_{\mathrm{D1}} + \sigma^{2}_{\mathrm{D2}} + \sigma^{2}_{\mathrm{ET}} + \sigma^{2}_{\mathrm{S}}}
\end{equation}
With:
\begin{itemize}
	\item $\sigma_{\mathrm{L}}$, timing uncertainty due to the laser pulse duration
	\item $\sigma_{\mathrm{D1}}$, time jitter of the start detector (photodiode)
	\item $\sigma_{\mathrm{D2}}$, time jitter of receive detector (SPAD)
	\item $\sigma_{\mathrm{ET}}$, time uncertainty of event timer
	\item $\sigma_{\mathrm{S}}$, uncertainties caused by the design of the satellite retroreflectors. These are not subject to the miniSLR design and are therefore not considered here. For most satellites, the satellite signature is very small compared to the other contributing factors. 
\end{itemize}

When evaluating equation \ref{eq:precision_terms}, one has to observe that timing uncertainties are given by different metrics, the most common being FWHM (full width half maximum), RMS (root mean square), or standard deviation $\sigma$ of a normal distribution. Assuming that all uncertainties are normal distributed (which is usually a reasonable approximation), it is possible to relate these quantities to each other with:

\begin{eqnarray}
	\sigma &=& 0.42 \times \mathrm{FWHM}   \\
	\sigma &\approx& \mathrm{RMS}    
\end{eqnarray}

\begin{table}
	\begin{tabular}{rrl}
		\toprule
		Laser              & \SI{210}{\pico\second}  & (Given as \SI{500}{\pico\second} FWHM) \\ 
		Start detector     & \SI{40}{\pico\second} & \\ 
		Receive detector   & \SI{150}{\pico\second} & (Worst case, probably better) \\ 
		Event timer        & \SI{9}{\pico\second} & \\ 
		\midrule
		Single shot precision $\sigma_{\mathrm{single}}$ & \SI{261}{\pico\second} &  (equivalent to \SI{39}{\milli\metre})\\  	
		\bottomrule
	\end{tabular} 
	\caption{Timing uncertainty budget for the miniSLR components. Single shot precision is calculated using equation \ref{eq:precision_terms}.}
	\label{tab:precision}
\end{table}

Using the specified values from the used components, an expected single shot precision of \SI{39}{\milli\metre} is derived (see Table \ref{tab:precision}). 

\subsubsection{Normal point precision}
\label{sec:NPT_precision_theo}
In post processing, individual range measurements are averaged into normal points (NPTs). Recommended normal point durations are given by the ILRS, and range from \SI{5}{\second} to \SI{300}{\second}, depending on satellite altitude and expected return strength. Assuming a purely statistical error distribution, the precision of a normal point $\sigma_{\mathrm{NPT}}$ with $N$ individual data points is given by

\begin{equation}\label{eq:precision_NPT}
	\sigma_{\mathrm{NTP}} = \frac{\sigma_\mathrm{single}}{\sqrt{N}}
\end{equation}

Given the single shot precision of \SI{39}{\milli\metre}, averaging about 1,500 data points in a normal point should yield a normal point precision of \SI{1}{\milli\metre}.

It must be pointed out that in reality the error distribution will not be purely statistical, but systematic errors also contribute. The main issue are drifts of timing delays (e.g. from photon detection to electrical signal), if they occur on the timescales of minutes. Drifts on longer timescales are eliminated by regular timing calibration. The amount of these contributions cannot be inferred from device specifications, but they are included in the experimental system validation (see Section \ref{sec:accuracy}).

\subsubsection{Return rates}
\label{sec:exp_return_rates}
The system is designed to always operate in single photon mode, i.e., for each outgoing laser pulse the detector should see no more than one photon. This avoids any systematic time shifts in the receive detector due to multiple photon signals. As the number of actual photons returning is following a Poisson distribution, this can be achieved by ensuring a mean return quota (received photons per outgoing pulse) of much less than one, ideally below 0.1.

On the other hand, the mean number of photons must no be too small in order to still produce a visible signal. The actual limit is hard to estimate and depends on background brightness, and a number of system specifications. For the described miniSLR set-up, the night time limit is around $5 \times 10^{-5}$ (equivalent to roughly \SI{2.5}{\hertz} return rate).

The expected return quotas for different targets, atmospheric conditions and measurement geometries can be calculated using the modified radar link equation from \cite{degnan1993millimeter}. It gives the mean number of expected photoelectrons per laser pulse as 
\begin{equation}
	n_{\mathrm{pe}}=
	\left(E_\mathrm{T}\frac{\lambda}{hc} \right)
	G_{\mathrm{t}}~
	\sigma_{\mathrm{ocs}}~
	\left(\frac{1}{4 \pi R^2}\right)^2
	A_{\mathrm{r}}~
	T^{2}_{a}~ 
	\eta_{\mathrm{t}}~
	\eta_{\mathrm{r}}~
	\eta_{\mathrm{d}}
	\label{eq:radar_equation}
\end{equation}
with:
\begin{itemize}
	\item $E_\mathrm{T}$, Laser pulse energy
	\item $\lambda$, Laser wavelength
	\item $G_{\mathrm{t}}$, Gain, a function of beam divergence and pointing stability
	\item $\sigma_{\mathrm{ocs}}$, Optical cross section of satellite reflector
	\item $\left(\frac{1}{4 \pi R^2}\right)^2$, Attenuation at distance $R$
	\item $A_{\mathrm{r}}$, Aperture of receive telescope
	\item $T^{2}_{\mathrm{a}}$, Atmospheric transmission
	\item $\eta_{\mathrm{t}}$, Efficiency of transmitter optics
	\item $\eta_{\mathrm{r}}$, Efficiency of receiver optics
	\item $\eta_{\mathrm{d}}$, Efficiency of detector
\end{itemize}

As some of the factors in equation \ref{eq:radar_equation} can only be estimated (e.g.\ beam pointing stability), or are subject to frequent changes (e.g.\ atmospheric conditions), the resulting numbers can be indicative only. A model based on this equation, but including elevation dependent atmospheric effects, has been developed and experimentally verified by \cite{tristan_master}. Using this model and the miniSLR specifications from Table \ref{tab:minislr_specs}, expected return quotas for a few important satellites have been estimated (see Table \ref{tab:return_quota_theo}). 

The numbers rapidly decrease with satellite altitude, owing to the $R^4$ factor in the link budget. While a very strong signal is expected from most LEO satellites, high satellites (especially Galileo) seem to be quite challenging. For low satellites, the return quota may even exceed the desired single photon maximum of 10\%, if indeed these theoretical values can be achieved. In this case, the beam steering could be used to slightly misalign the beam, to reduce the return quota.

It should be pointed out that the optical cross sections used here are lower than theoretical values from \cite{arnold_cross_sections}. Also, some of the miniSLR specifications are rather cautious, e.g.\ the receiver efficiency of $10\%$. It is thus conceivable that higher return rates than calculated here are possible in reality, especially under favourable atmospheric conditions.

In Section \ref{sec:return_rates_meas}, some experimentally measured return rates are compared with the values estimated here.

\begin{table}
	\begin{tabular}{rrrrr}
		\toprule
		Satellite   & NPT Duration       & Optical cross section & Return quota & returns / NPT \\
		\midrule
		Grace-FO  & \SI{5}{\second}   & \SI{0.6e6}{\metre^2}  & 70\% & 175,000\\
		Ajisai    & \SI{30}{\second}  & \SI{6.1e6}{\metre^2} & 18\% & 270,000\\
		Stella    & \SI{30}{\second}  & \SI{0.1e6}{\metre^2} & 3.7\%  & 55,000 \\
		Lares     & \SI{30}{\second}  & \SI{0.28e6}{\metre^2} & 1\%  & 15,000  \\
		Lageos    & \SI{120}{\second} & \SI{1.24e6}{\metre^2} & 0.02\% & 1,200\\
		Etalon    & \SI{300}{\second} &	\SI{23e6}{\metre^2}   & 0.004\% & 600\\
		Galileo   & \SI{300}{\second} &	\SI{3.1e6}{\metre^2} & 0.0003\% & 45\\
		\bottomrule
	\end{tabular}	
	\caption{Satellite optical cross sections, return quotas and number of returns per normal point (NPT) expected for different satellites, based on the model referenced in Section \ref{sec:exp_return_rates}.}
	\label{tab:return_quota_theo}
\end{table}

\section{Validation results and discussion}
\label{sec:results}

\subsection{Data census and processing}
\label{sec:data}

The results presented here are based on observations made between November 2022 and April 2023. 

\begin{figure}[p]
	\centering
	\includegraphics[width=0.8\textwidth]{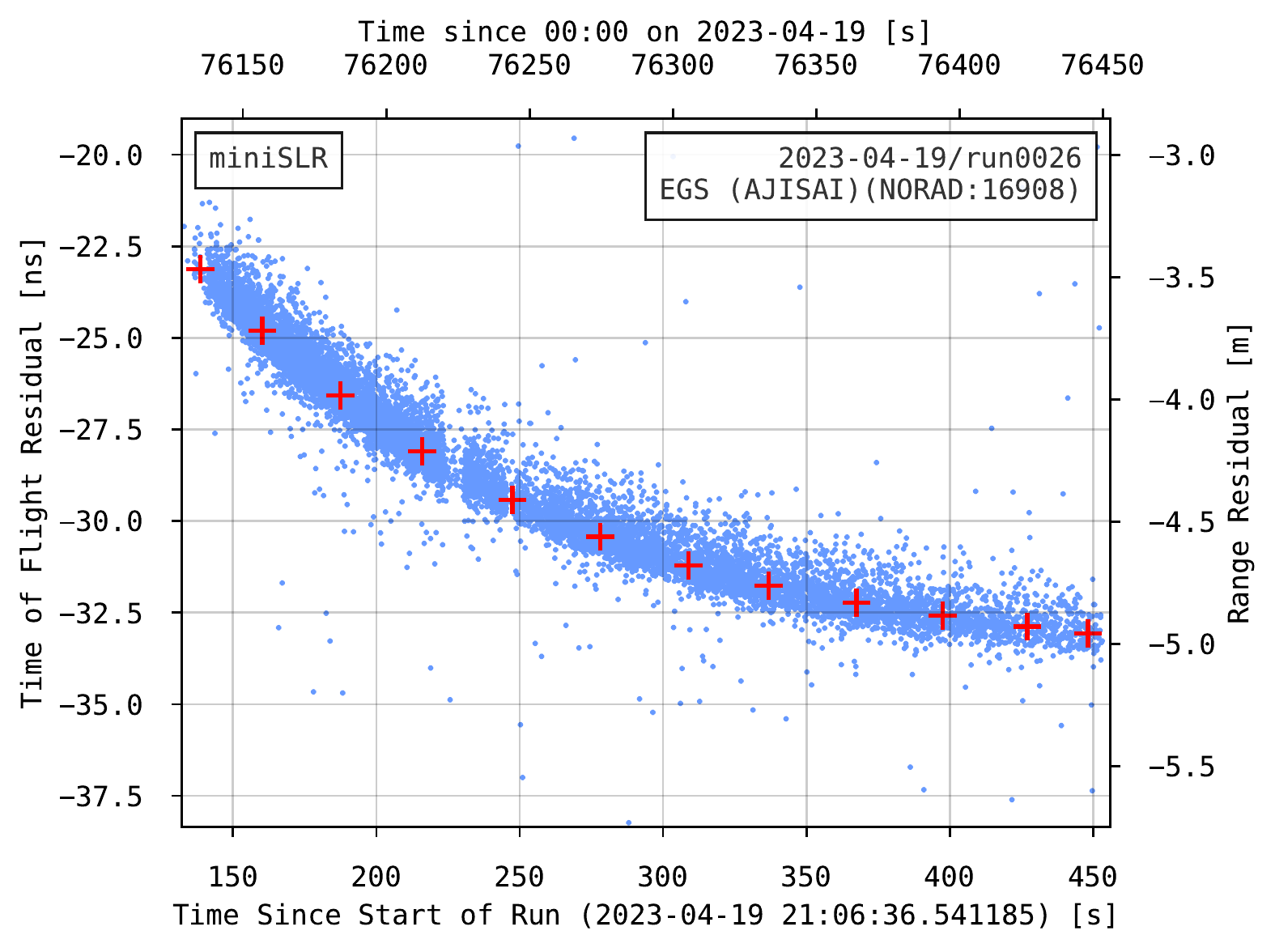}
	\caption{Ranging plot of Ajisai. For clarity, only every 100th data point is shown. Normal points are marked in red. They contain between 40,000 and 200,000 individual ranges.}
	\label{fig:ranging_ajisai}
\end{figure}

\begin{figure}[p]
	\centering
	\includegraphics[width=0.8\textwidth]{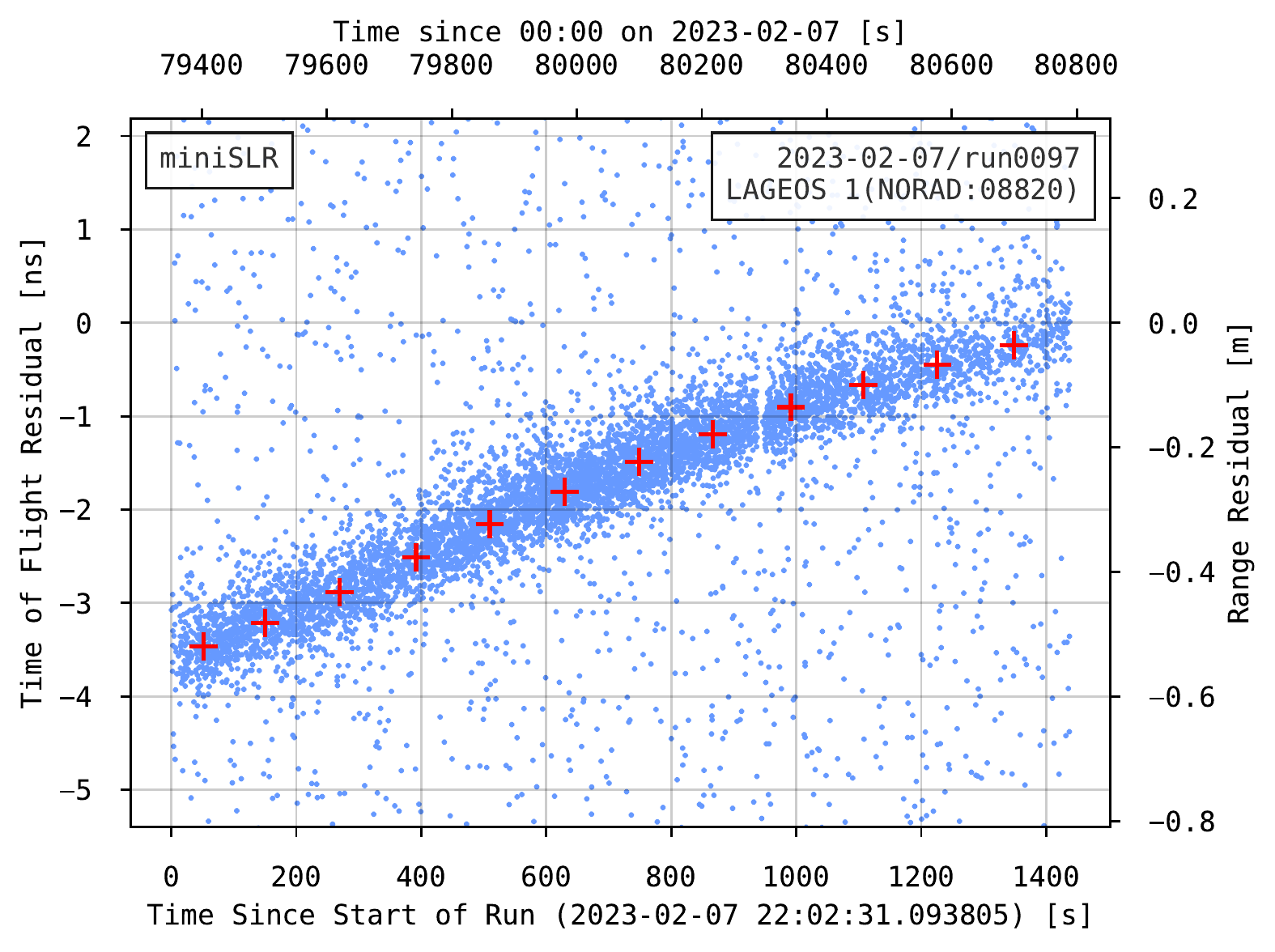}
	\caption{Ranging plot of Lageos 1. For clarity, only every 3rd data point is shown. Normal points contain about 2,000 ranges each.}
	\label{fig:ranging_lageos}
\end{figure}

\begin{figure}[t]
	\centering
	\includegraphics[width=0.8\textwidth]{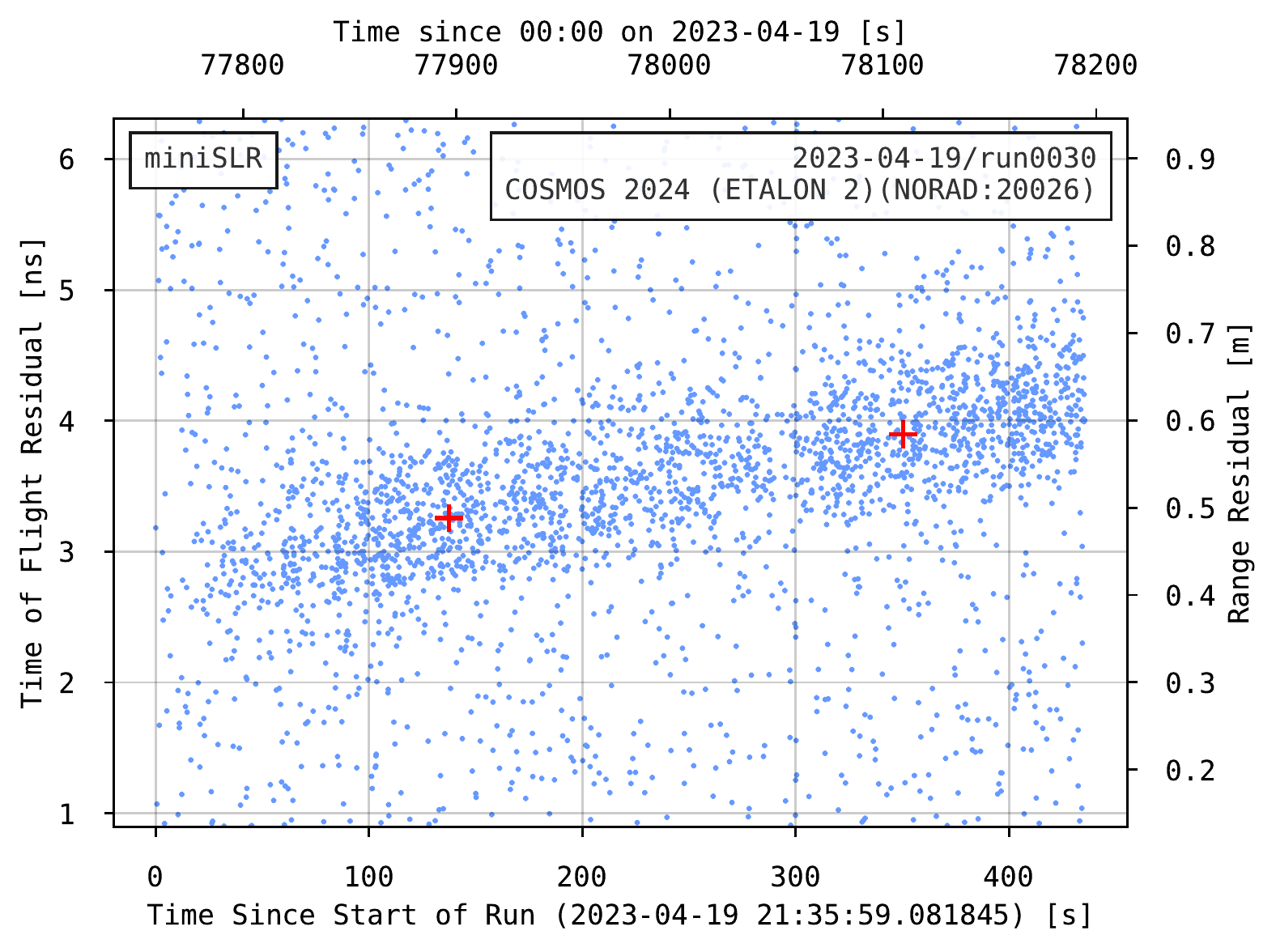}
	\caption{Ranging plot of Etalon 2. Normal points contain around 800 ranges each.}
	\label{fig:ranging_etalon}
\end{figure}

Figures \ref{fig:ranging_ajisai} to \ref{fig:ranging_etalon} show some typical examples of measurements taken. Post-processing of the data is done manually using OOOS. Data filtering and normal point generation are implemented according to the algorithm developed by the ILRS (\cite{npt_algorithm}). A rejection interval of 2.5 times the RMS is used, as recommended for single photon systems. The normal points are indicated by red crosses in the plot. Summary measurement reports are generated in CRD (consolidated ranging data) format, and used for further analysis.

\subsection{Tracking accuracy}
Accurate satellite tracking is fundamental to a productive SLR operation, but at the same time challenging for a relatively small mount. To allow for blind tracking (without visual acquisition), the tracking accuracy should be not much worse than the laser beam divergence of 10 arcsec ($\approx \SI{50}{\micro rad}$). 

For the generation of pointing models, about 50-70 stars are recorded automatically. The process is slightly complicated by the fact that the mount cannot move the full 360° in azimuth, and has to use both pier sides (i.e.\ elevations above 90°) to cover the full sky. For each star, the offset from the camera target point is recorded. All offsets are subsequently fitted by an analytical model adapted from \cite{tpoint}.

With this, a pointing accuracy of better than 10 arcsec is achieved both on stars as well as on satellites with accurate predictions, including fast LEO satellites. Ranging with blind tracking has been shown successfully in a number of cases. Unfortunately, the pointing accuracy quickly deteriorates, and blind tracking usually becomes impossible after a few weeks. The reason for this is not yet determined, but it may be due either to instability in the mounting (on a gravel bed on the roof of a six-storey building), or some thermal effects in the mechanical mountings of the optical system.

For the current study, most passes have been recorded at night, with visual guidance. Closed loop tracking is performed automatically by the software if the target is visible. A few passes have been recorded with blind tracking, or partial blind tracking (visual acquisition before entering the earth shadow).

Daylight tracking has been attempted once, but without success. Detector rates, however, seemed to be at a manageable level. Calibration records could be recorded without issues. The failure to see satellite returns is believed to be due to insufficient pointing accuracy. 

Improving the pointing stability will be an important task in the further development of the miniSLR, to enable blind tracking at day and night.

\subsection{Return rates}
\label{sec:return_rates_meas}

\begin{figure}[p]
	\centering
	\includegraphics[width=0.8\textwidth]{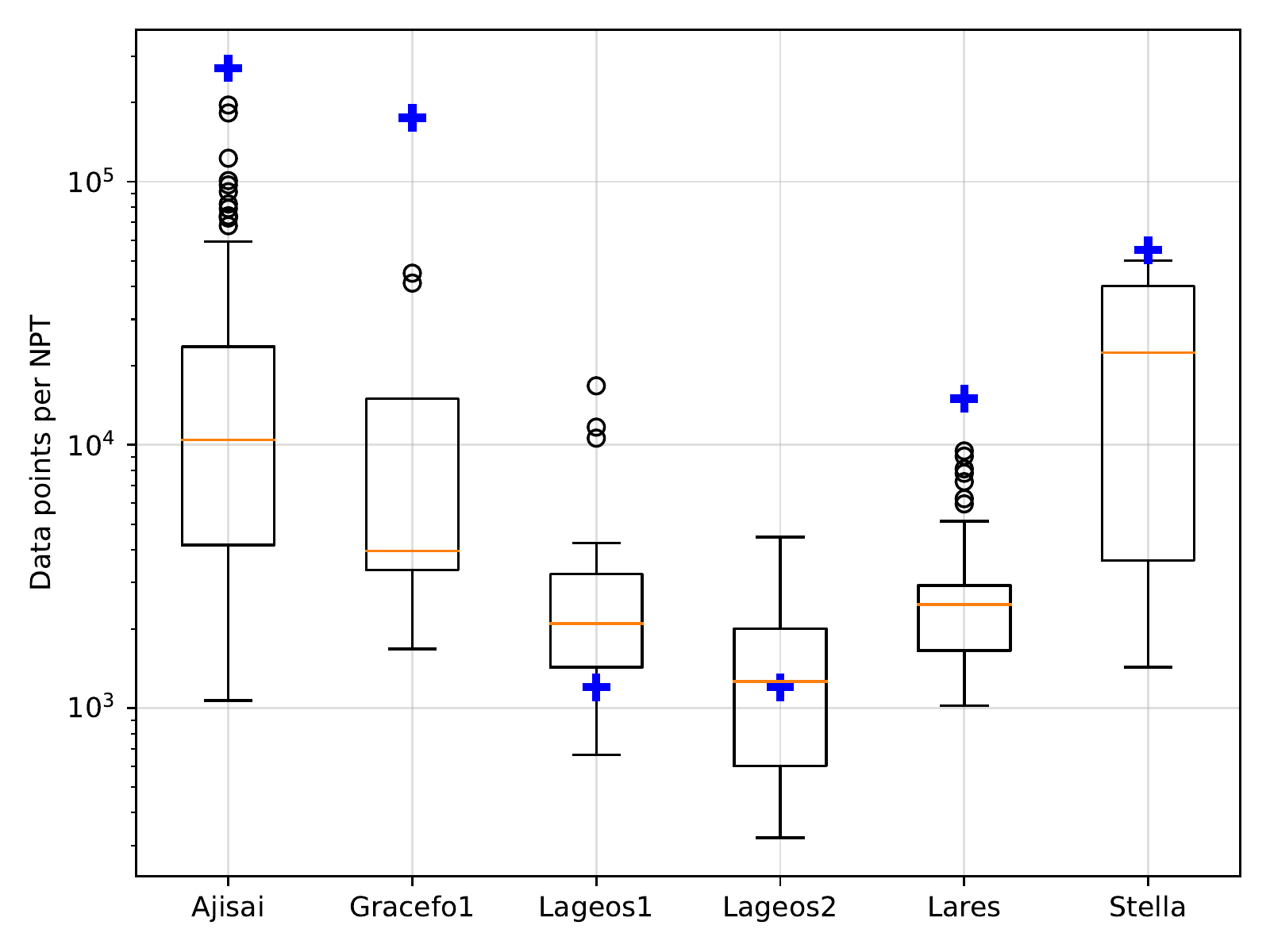}
	\caption{Measured return rates, given in data points per normal point (NPT). Normal point durations are \SI{5}{\second} for Grace-FO, \SI{30}{\second} for Ajisai, Lares and Stella, and \SI{120}{\second} for the Lageos satellites. The boxes show the range from first to third quartile of the return numbers, the horizontal line denotes the median. Outliers beyond twice the inter-quartile range are shown as individual circles. Blue crosses mark the theoretical expectation from Table \ref{tab:return_quota_theo}.}
	\label{fig:return_rates_meas}
\end{figure}

\begin{figure}[p]
	\centering
	\includegraphics[width=0.8\textwidth]{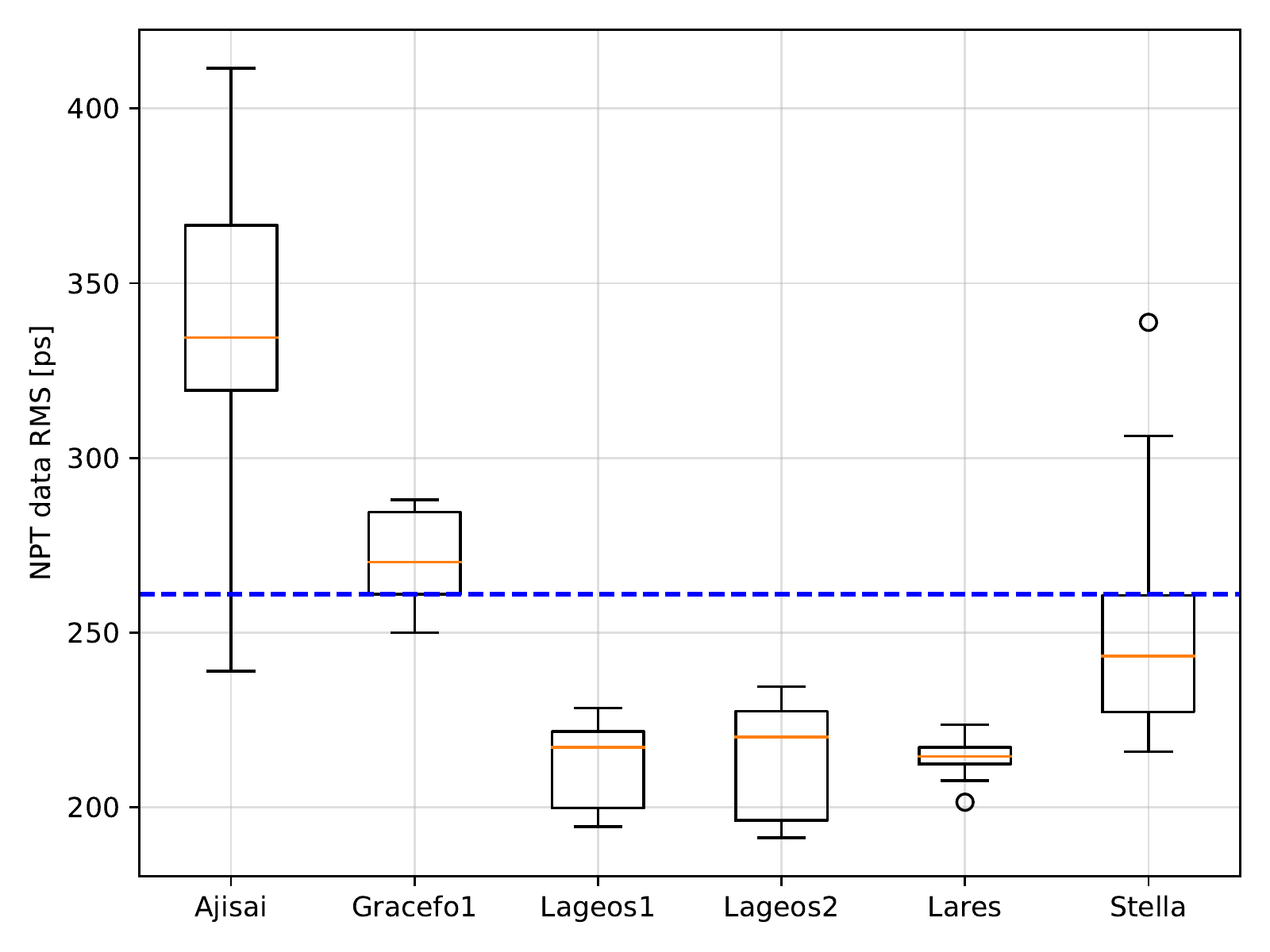}
	\caption{Measured RMS of data within a normal point (NPT). The theoretical expectation for this value is given as \SI{261}{\pico\second} (blue dashed line, see Table \ref{tab:precision}). See caption of Figure \ref{fig:return_rates_meas} for further explanation.}
	\label{fig:RMS_meas}
\end{figure}

Figure \ref{fig:return_rates_meas} shows measured return numbers per normal point for some satellites. By and large, the measured numbers correspond roughly with the theoretical expectations from Section \ref{sec:exp_return_rates} (blue crosses). As expected from the modelling, large spreads between high and low data yields exist. These can be attributed to differences in tracking geometry, elevation angle, atmospheric transmission (local thin clouds), and changes in tracking accuracy.

Except for Lageos, the experimental values are somewhat lower than the calculated values, which are already at the low end of theoretical expectations. This may indicate that system losses are higher than assumed, and higher return rates may be achieved e.g.\ by better alignment of the optics.

\subsection{Precision}
\label{sec:precision}

During the post-processing, the single shot RMS is calculated for each normal point. In calibration runs, the values are typically between \SI{210}{\pico\second} and \SI{230}{\pico\second}. In satellite tracks, the RMS depends slightly on the strength of the signal, probably due to imperfections in the data filtering. Figure \ref{fig:RMS_meas} shows the normal point RMS values for a selection of satellites. Median values range from \SI{220}{\pico\second} to \SI{330}{\pico\second}. This is well compatible with the theoretical expectation of \SI{261}{\pico\second} / \SI{39}{\milli\metre} (see Section \ref{sec:precision_theo}).

As can be seen from Figure \ref{fig:return_rates_meas}, the expected minimum of 1,500 data points per normal point is often achieved. For some Lageos points, and usually for high satellites (such as Etalon-2, shown in Figure \ref{fig:ranging_etalon}), it can be lower. 

A quality cut has been applied at 300, i.e. NPTs with less than 300 data points are discarded. Such low data normal points may occur at e.g.\ beginning or end of measurements, during interruptions due to aircraft warnings, from imperfect tracking, or scattered clouds. Ideally, one would like to increase the quality cut to 1,500 points, to achieve the envisaged averaging effect (see Section \ref{sec:NPT_precision_theo}), however this would have eliminated too much data in the present study. With a higher quality cut, a slightly improved normal point precision may be achieved.

\subsection{Accuracy}
\label{sec:accuracy}

To obtain a realistic estimation of the system performance, the data has kindly been analysed by Toshimichi Otsubo from Hitotsubashi University, using his rapid quality control software (\cite{otsubo_quality_control}). It generates global fits for the orbits of all considered satellites, based on data from all SLR stations that have uploaded measurements for the time period in question. Thus, the analysis provides a measure of one station's accuracy, relative to all other stations in the network.

The most relevant results of this analysis are:
\begin{itemize}
	\item Station coordinates 
	\item Station range bias (adjusted once for each station)
	\item Pass range bias (adjusted for every pass)
	\item Normal point precision, estimated from scatter of normal points around fit.
\end{itemize}

For the analysis of the miniSLR accuracy, only data taken during 5 nights from February 7 to 13, 2023, is considered. Five satellites are included in the analysis: Lageos 1 and 2, Ajisai, Stella and Lares. The data comprises of 15 passes with a total of 163 normal points of these satellites. Based on this limited dataset, a first estimation of the experimental performance of the system is performed. 

\begin{table}
	\begin{tabular}{crr}
		\toprule
		Coordinate & Local survey     & SLR analysis \\
		\midrule
		x          & 4,160,755.242 m  & 4,160,755.135 m \\
		y		   &   666,638.631 m  &   666,638.658 m \\
		z		   & 4,772,593.195 m  & 4,772,593.327 m \\
		\bottomrule
	\end{tabular}	
	\caption{Station coordinates in ITRF2014, as measured by a local surveyor, and by SLR data from February 2023.}
	\label{tab:station_coordinates}
\end{table}

Coordinates of the station invariant point (intersection of the two mount axes) have been measured by a surveyor in a local datum, and transformed to ITRF2014 Cartesian coordinates. They agree to the coordinates from SLR data to within \SI{20}{\centi\metre} (see Table \ref{tab:station_coordinates}). The reason for this rather large deviation may be inaccuracies in the conversion of the local datum into ITRF2014, or insufficient SLR data for a very accurate position esimate. 

The station range bias is fitted with \SI{3.4}{\centi\metre}. While already encouragingly small, this number is still larger than expected from system specifications. Possible reasons could be a wrongly measured distance to the local calibration target, or systematic biases between calibration and actual satellite measurements (e.g.\ by the attenuator).

\begin{figure}[pt]
	\centering
	\includegraphics[width=0.8\textwidth]{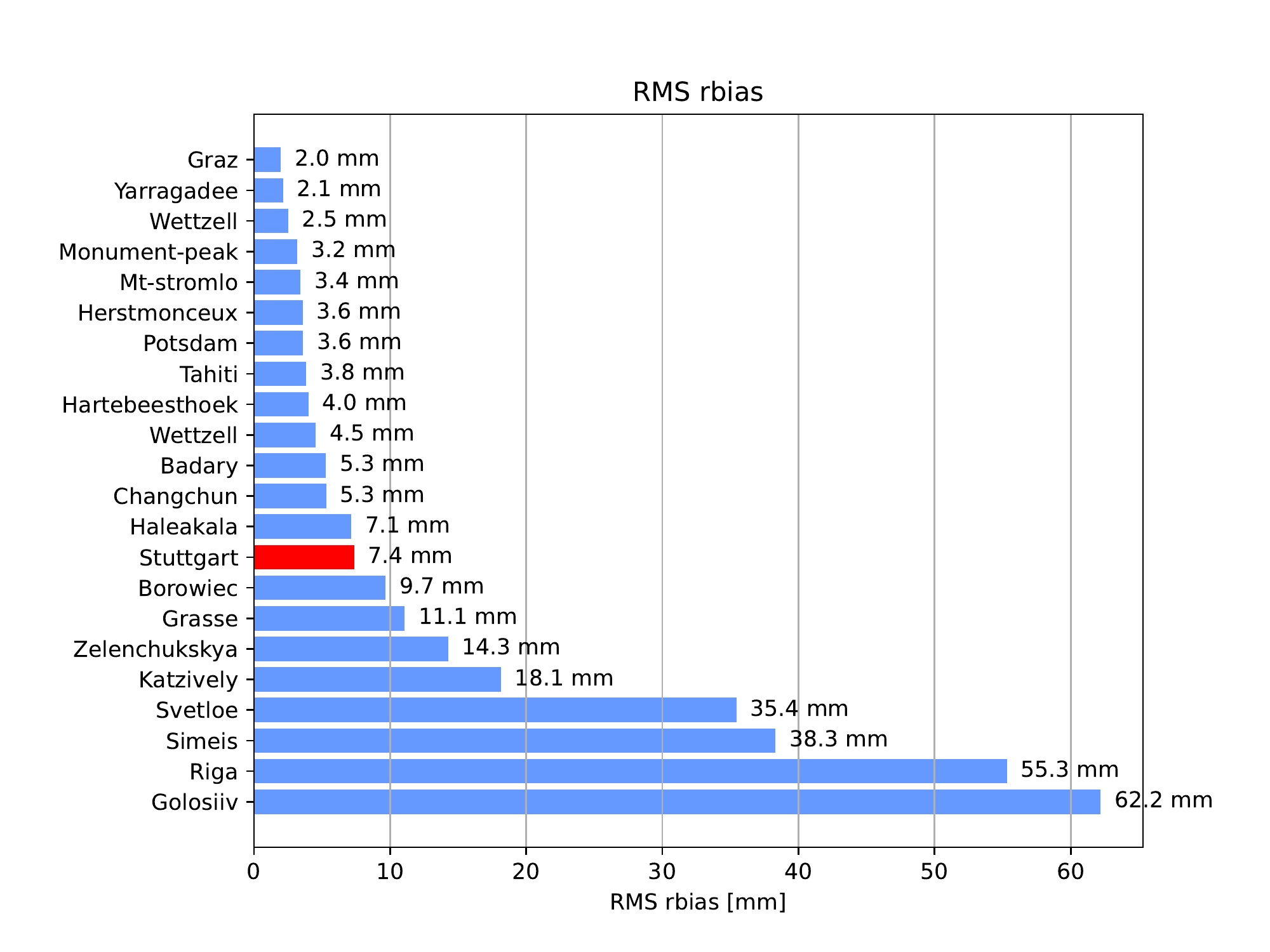}
	\caption{RMS of pass range biases according to data analysis described in Section \ref{sec:accuracy}. It displays the changes in the pass-to-pass range bias offsets applied to match the global orbital fits.}
	\label{fig:RMS_rbias}
\end{figure}

\begin{figure}[pt]
	\centering
	\includegraphics[width=0.8\textwidth]{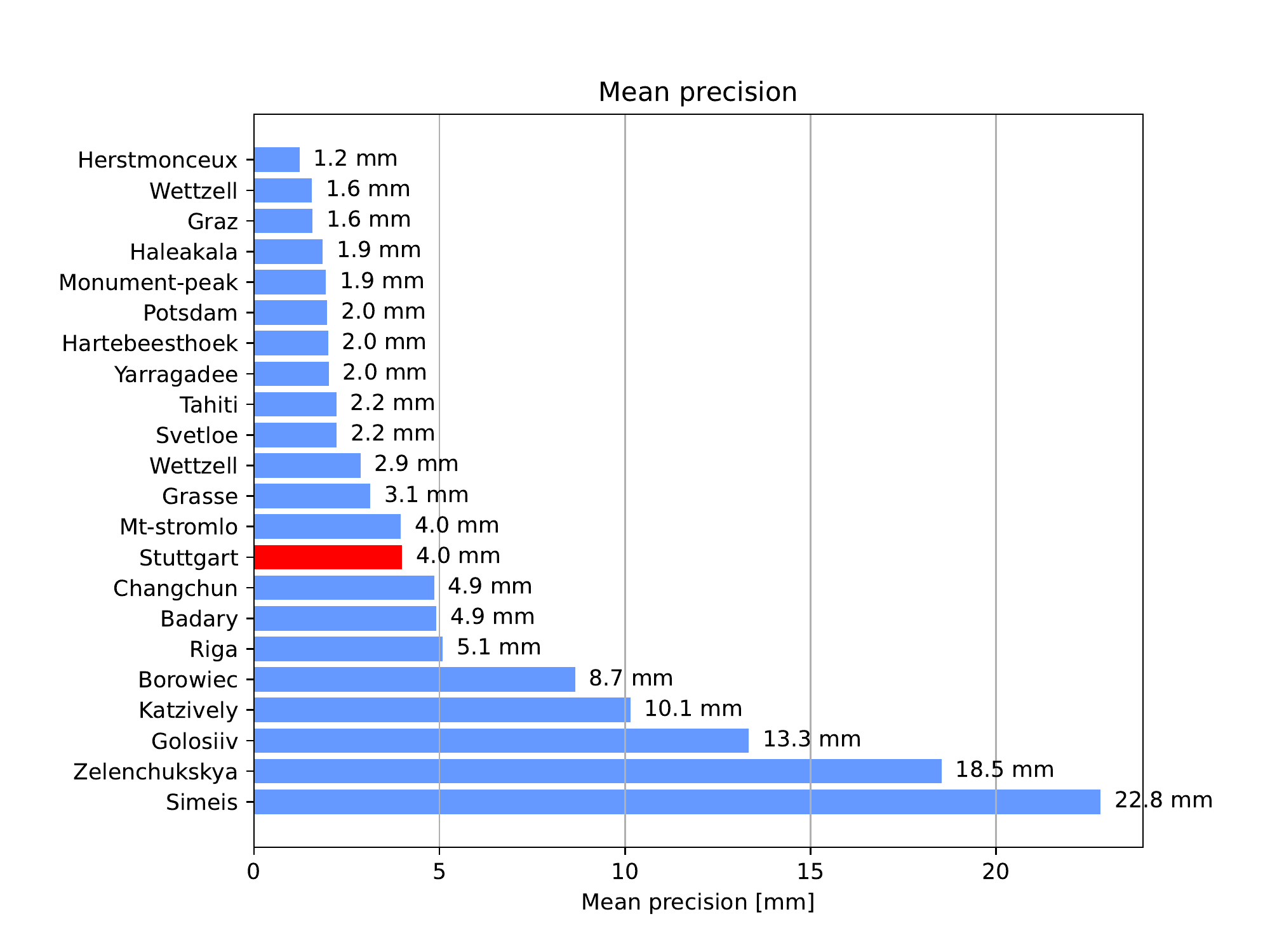}
	\caption{Mean precision according to data analysis described in Section \ref{sec:accuracy}. The mean precision is given as scatter of normal points around the fitted orbit, after application of a constant range bias, and a pass-dependent range and time bias.}
	\label{fig:NPT_precision}
\end{figure}

The pass-by-pass variation of the range bias is found to have an RMS of \SI{7.4}{\milli\metre}. This can be compared to the pass range bias RMS of the other stations that have supplied data in the same timespan for the same satellites (see Figure \ref{fig:RMS_rbias}). While the best stations in the network, like Graz, Yarragadee or Wettzell, achieve values below \SI{3}{\milli\metre}, other station are at the same order of magnitude as the Stuttgart miniSLR, or worse. 

The normal point precision, i.e.\ the scatter of normal points around the fitted (and bias-corrected) orbits, averages to \SI{4}{\milli\metre}. Again, the comparison with other station shows that the miniSLR achieves a satisfactory performance (Figure \ref{fig:NPT_precision}).

While the low number of data points limits the statistical significance of these results, they are a good indication of the possible performance of the system. It seems fair to claim that the miniSLR can indeed reach a similar accuracy as conventional stations, and thus be valuable tool for geodesy and other SLR applications. Long-term stability and more statistical significance of the results will require more data, which will be collected and supplied to the ILRS in the future.

\section{Conclusion and Outlook}
\label{sec:conclusion}

In the scope of the work described here, a fully functional prototype of a minimal SLR system has been constructed, commissioned and tested. The validation was done on existing ILRS supported satellites at all relevant altitudes and with different retroreflector configurations. The results indicate that ranging to most relevant targets can be performed with an accuracy comparable to existing, conventional SLR systems. 

Compared to state-of-the-art SLR systems, the miniSLR currently still lacks the possibility to consistently range to Galileo satellites, and to reliably perform blind ranging (without visual acquisition). Also, daylight ranging has not yet been demonstrated. 

The issues with blind tracking can probably be solved with a more rigid mechanical construction of the optical bench, or even just with a more suitable operating location. Presumably, this would also enable daylight ranging. While the effect of such improvements is yet to be shown, the problems seem not to be immanent to the minimal SLR concept.

The issue of ranging to Galileo targets, on the other hand, is indeed connected to the small size of the receiver aperture. Both theoretical estimates as well as experimental results seem to indicate that despite the rather high laser power of \SI{5}{\watt}, not enough photons are received to reliably detect returns from these satellites. It may be possible to achieve a better sensitivity by improving the alignment of the optics, but this is uncertain. Other options, like a more powerful laser, can be considered as well, but would require a significant engineering effort.

Thus, the current version of the miniSLR seems mainly suitable for the following applications:
\begin{itemize}
	\item Geodetic measurements, especially at remote locations currently not well covered by the existing SLR network
	\item Supporting high-performance stations, relieving them of some of the daily tracking load
	\item Mission support for LEO satellites
	\item Conjunction assessment, if at least one object is equipped with retroreflector
	\item Studies and experiments requiring a flexible SLR ground station
\end{itemize}

The Stuttgart miniSLR will continue to deliver data to the ILRS for further validation. As part of a research project, it will also be equipped with polarization optics to test the feasibility of satellite identification through polarizing retroreflectors (\cite{nils_pola}). Additionally, it will be used to participate in laser ranging research projects.

In parallel, a commercial version of the miniSLR will be designed and constructed by DiGOS Potsdam GmbH. Using a different mount, an improved mechanical set-up and a number of smaller modifications, it is expected to achieve an even better performance than the prototype described in this paper.

\section*{Declarations}

\subsection*{Acknowledgements}
The authors would like to acknowledge contributions to this project by 
\begin{itemize}
	\item former team member Ewan Schafer,
	\item former team member Paul Wagner,
	\item Robin Neumann and Luis Gentner, who supported the construction and data taking,
	\item the institute's mechanical, electronic and IT department for their continued effort in supporting this project,
	\item Prof.~Toshimichi Otsubo from Hitotsubashi University, Tokyo, for his continued support and encouragement. 
\end{itemize}

\subsection*{Supplementary information}
The miniSLR design is partly patented. The name "miniSLR" is a registered trademark.

\subsection*{Funding}
The work described here was mainly funded by DLR as part of a technology transfer project.

\subsection*{Conflict of interest}
DH declares that he is involved in the commercial distribution of the miniSLR at DiGOS Potsdam GmbH and may profit from sales of the system.

\subsection*{Availability of data}
The normal point data generated and analysed for this paper is available on the EDC website (\cite{EDC_web}). Raw data is available from the corresponding author on reasonable request.

\subsection*{Code availability}
The code used to process and analyse the data is part of the software package OOOS, which is licensed under GPLv3. It is available from the corresponding author on reasonable request.

\subsection*{Author contributions}
DH co-developed the concept of the miniSLR, lead the technical development and wrote the paper. FN worked on the technical implementation of the miniSLR and conducted measurements. TM conducted measurements, calculated theoretical return rates, and supported the software development. WR co-developed the concept and vision of the miniSLR and had the administrative lead of the project.


\bibliography{sn-bibliography}

\end{document}